\documentclass[a4paper,english]{article}
\usepackage{amsmath,amssymb}
\usepackage{bm}
\usepackage{graphicx,color}
\usepackage{fancybox}

\newtheorem{theorem}{Theorem}[section]

\newtheorem{idea memo}[theorem]{Idea Memo}
\newtheorem{lemma}[theorem]{Lemma}

\newtheorem{proposition}[theorem]{Proposition}

\input{tcilatex}

\newcommand{\EX}[1]{\text{Exp\ }(#1)}
\newcommand{\naiseki}[2]{\langle #1,#2\rangle}

\newcommand{\limn}{\lim_{n\to \infty}}
\newcommand{\kakko}[1]{\left (#1\right )}
\newcommand{\eqa}[1]{
\begin{align*}
#1
\end{align*}}

\newcommand{\dom}{\mathcal{O}}
\newcommand{\dis}{\displaystyle}

\newcommand{\MG}{C_c^{\infty}(M,G)}
\newcommand{\Mg}{C_c^{\infty}(M,\mathfrak{g})}

\newcommand{\supp}[1]{\text{supp}(#1)}

\newcommand{\pf}{\textbf{Proof.}}
\DeclareMathOperator*{\UU}{\cup}
\DeclareMathOperator*{\CAP}{\bigcap}
\DeclareMathOperator*{\OP}{\bigoplus}

\begin{document}
\title{On the local structure of the representation of a local gauge group}
\author{Hiroshi Ando}
\date{Research Institute for Mathematical Sciences, Kyoto University, Japan}
\maketitle
\begin{center}
\textbf{Abstract.}
\end{center}
\small We discuss the local structure of the net $\mathcal{O}\mapsto \mathcal{M}(\mathcal{O})''$\ of von Neumann algebras generated by a representation of a local gauge group $\MG$.\ Our discussion is independent of the singularity of spectral measures, which has been discussed by many authors since the pioneering work of Gelfand-Graev-Ver\v{s}ic.\ We show that, for type (S) operators\ $U_{A,b}$, second quantized operators with some twists,\ the commutativity only with those $U(\psi)$ is sufficient for the triviality of them, where $\psi$ belongs to an arbitrary (small) neighborhood of constant function 1. Some properties of 1-cocycles for the representation $V:\psi \mapsto \text{Ad}_{\psi}$ are also discussed.\normalsize      
\section{Introduction}
In this paper we consider a representation $\psi \mapsto U(\psi)$\ of a local gauge group\ $C_c^{\infty}(M,G)$\ defined in the Boson Fock space\ $\Gamma(H)$, where the Hilbert space $H$ is the completion of the space of connection 1-forms on a manifold $M$:\ 
\[U(\psi)\text{Exp }\omega :=e^{-\frac{1}{2}||\beta(\psi)||^2-\naiseki{V(\psi)\omega}{\beta (\psi)}}
\text{Exp}\kakko{V(\psi )\omega +\beta (\psi)},\ \omega \in H.\]
where, $V(\psi):=\text{Ad}_{\psi}, \beta(\psi):=d\psi \cdot \psi^{-1}.$\\
\ \ Roughly speaking, we consider the second quantization of the action of $\psi \in \MG$\ on $H$\ defined by
\[\omega \mapsto \psi \omega \psi^{-1}+d\psi \cdot \psi^{-1},\ \omega \in H.\]
The study of this representation seems to have started around 70's  by Gelfand-Graev-Ver\v{s}ic\cite{GelV2}\ (for $SL(2,R),R=$function space) and by Ismagilov \cite{Ism}\ (for $G=SU(2)$).\ The present form of the representation first appeared in Gelfand-Graev-Ver\v{s}ic\cite{Gel77}, in which they proved the irreducibility for $\dim(M)\ge 2$\ and semisimple compact $G$.\ Although the proof was elegant, it contained some gaps. Later they proved the irreducibility for $\dim(M)\ge 4$\ in \cite{Gel80}.\ Unfortunately it still contained a mistake, as was pointed out by Wallach\cite{Wal}. On the other hand, Albeverio-H\o egh Krohn-Testard\cite{Alb81} proved the irreducibility for $\dim(M)\ge 3$\ and $\dim (M)=2$\ with some conditions on the size of root vectors.\ Later Wallach proved in \cite{Wal} the irreducibility for $\dim(M)\ge 3$\ and $\dim (M)=2$\ under weaker conditions than \cite{Alb81}. The $\dim(M)=2$\ case has not been completely settled yet.\ For the $\dim(M)=1$\ case, Albeverio-H\o egh Krohn-Testard proved that for $M=S^1$\ the representation is reducible and in fact constitutes a type III factor \cite{Alb81}.\ In the '90s Driver-Hall proved that there is no such $\Omega \neq 0 \in \Gamma(H)$\ that is invariant under all $U(\psi)$'s.\ Recently Y.Shimada\cite{Shimada} proved the irreducibility for all compact $M$ with arbitrary $\dim (M)$. He used the technique of Fock expansion, which is a fundamental tool of White Noise Analysis.\ However, there were some mistakes and the proof was not complete\footnote{Particularly he assumed that $\beta(e^{\varphi})=d\varphi$\ for any $\varphi \in \Mg$, which is true only when $\varphi $\ takes values in some abelian subgroup of $G$. In the general case we must compute the derivative of the exponential mapping carefully.\ cf,\ \cite{Grab}.
}, as remarked by T. Hasebe.\  
Almost all of the studies (except Shimada's) were based on the analysis of the disjointness properties between two spectral (in fact Gaussian ) measures related to the representations of the abelian subgroup $\EX{C_c^{\infty}(M,\mathfrak{h})}\subset C_c^{\infty}(M,G).$\footnote{For the details of this method, see\ \cite{Gel77,Alb}}\ However, there seems to be no attmept to study the local structure of the representation.\ Therefore, we study in this paper the algebraic structures of the type (S) representation with its localization aspects in focus,  according to the suggestion by I. Ojima. 
\ We focus on the von Neumann subalgebra $\mathcal{M}(\mathcal{O})''$\ generated by the operators $U(\psi)$ whose supports are contained in $\dom.$\ The structure of the type (S) operators(see $\S 2$)
\[U_{A,b,c}\text{Exp }x:=c\cdot e^{-\frac{1}{2}||b||^2-\naiseki{Ax}{b}}\text{Exp }(Ax+b),\]
looks like the symplectic structure of Weyl unitaries\ $W(h)$, which are related to the von Neumann algebra of free Bose fields\cite{Araki1}. However, we show that there is a sharp difference between free field algebras and the algebra of the gauge group representation.\ Namely, we prove\\
\ \ \textbf{Theorem.}\\
\textit{Let us define $\mathcal{A}:=\textrm{Lin}\{U_{A,b,c};(A,b,c)\in \mathcal{U}(H)\times H\times \mathbb{T}\}, \mathcal{A}(\mathcal{O}):=\text{Lin}\{U_{A,b};A|_{H(\mathcal{O'})}=\text{Id}_{H(\mathcal{O'})},\ AH(\dom)\subset H(\dom),\ \text{Int}(\text{supp}(b)\cap \mathcal{O'})=\phi\ \}.$}
\textit{Let $N_0$ be a neighborhood of $1\in \MG$.\ Then for any open subset \ $\mathcal{O}\subset M$, we have}\\
(1)\ \textit{$\mathcal{M}(\mathcal{O})'\cap \mathcal{A}=\mathcal{A}(\mathcal{O'}).\ (\dom':=M\backslash \dom.)$\ In particular,
$\mathcal{M}'\cap \mathcal{A}=\mathbb{C}1.\ \mathcal{M}:=\mathcal{M}(M).$\ Furthermore, it holds that $U(N_0)'\cap \mathcal{A}=\mathbb{C}1.$}\\
(2)\ \textit{The net $\mathcal{O}\mapsto \mathcal{M}(\mathcal{O})''$\ satisfies} 
\[\begin{cases}
isotony:\ \mathcal{O}_1\subset \mathcal{O}_2\Rightarrow \mathcal{M}(\mathcal{O}_1)''\subset \mathcal{M}(\mathcal{O}_2)''.\\
locality:\ \mathcal{O}_1\subset \mathcal{O}_2'\Rightarrow \mathcal{M}(\mathcal{O}_1)''\subset \mathcal{M}(\mathcal{O}_2)'.\\
\dis additivity:\ M=\UU_{i}\mathcal{O}_i\Rightarrow \mathcal{M}''=\kakko{\UU_i \mathcal{M}(\mathcal{O}_i)}''.
\end{cases}\]
\textit{However, the Fock vacuum $\Omega :=\EX{0}$\ is not cyclic for local subalgebras $\mathcal{M}(\mathcal{O})'' $ with a proper subset $\dom \subsetneq M$.\ Furthermore, if the representation is irreducible, then $\Omega$ is not separating for $\mathcal{M}(\mathcal{O})''$.}\\
\ \ In particular, there are no clear modular-symplectic structure nor Reeh-Schlieder property for the net.\  The most important point in our proof is that there 
is a sharp difference between the behaviors of $V(\psi)\ : \omega \mapsto \text{Ad}_{\psi}\omega $\ and of  $\beta(\psi)=d\psi \cdot \psi^{-1}$. The latter is regarded as a 1-cocyle for the former\ ($\S 2, \S3$).\ The difference is manifest when we consider the infinitesimal gauge transformations\ ($\S 4$). We show through the proof of the above theorem that if type (S) operators commute with every $U(\psi)$ for $\psi $\ belonging to the member of arbitrary small neighborhood $N_0$ of constant function $1\in \MG$, then it is a scalar operator. Therefore even if the representation is reducible, the commutant is very small.
\section{Preliminaries}
\subsection{Boson Fock space and type (S) representations}
In this section, we summarize some well-known background materials relevant to our discussion. For the proof of the facts stated in this section, see e.g. Guichardet\cite{Gui}, Albeverio, et al\cite{Alb}.
\subsubsection{Operators of type (S)}
We describe the algebraic structure of type (S) operators, which constitute a weakly dense *-subalgebra of bounded operators in the Boson Fock space.\ Let $H$ be a complex Hilbert space,\footnote{
We define the inner product to be linear in the left variable and anti-linear in the right variable.},\ and  $\dis \Gamma(H)$\ its Boson (Symmetric) Fock space:
\[\Gamma(H):=\OP_{n\geq 0}H^{\hat{\otimes}n}.\]
($\hat{\otimes}$\ means a symmetric tensor product).\ In $\Gamma (H)$, the set of exponential vectors 
\[\left \{\EX{h}=\kakko{1,\ h,\ \dfrac{h^{\hat{\otimes}2}}{\sqrt{2!}},\cdots ,\ \dfrac{h^{\hat{\otimes}n}}{\sqrt{n!}},\ \cdots }\in \Gamma(H);\ h\in H\right \}\]
is  linearly independent and is total in $\Gamma(H)$\cite{Gui}.\ Consider the subset $\mathcal{S}=\{\lambda \text{Exp }x;x\in H,\lambda \in \mathbb{C}\}$ of $\Gamma(H)$.\\
A unitary operator $U\in \mathbb{B}(\Gamma(H))$ is called an operator of type (S) if it preserves the subset $\mathcal{S}$:  $U\mathcal{S}=\mathcal{S}$.\ The set of such operators is completely determined \cite{Gui}:\\
\ \ Let $\mathcal{U}(H)$ be the group of unitary operators in $H$,\ $\mathbb{T}=\{\lambda \in \mathbb{C};|\lambda |=1\}$ be the 1-dimensional torus.\ For $\ A\in \mathcal{U}(H),\ b\in H,\ c\in \mathbb{T}$, it is easy to see that the operators $U_{A,b,c}$ defined by
\[U_{A,b,c}\text{Exp }x:=c\cdot e^{-\frac{1}{2}||b||^2-\naiseki{Ax}{b}}\text{Exp }(Ax+b),\]
are of type (S).\footnote{The above definition is well-defined because of the independence and totality of exponential vectors. The unitarity can be verified by a straightforward calculation.} Moreover, the converse is also true.\ Namely, 
\begin{theorem}{\cite{Gui}}
\textit{All operators of type (S) are uniquely written as }$U_{A,b,c}$\textit{ for some }$A,b,c$.\ \textit{Moreover, due to the relations}
\[U_{A,b,c}U_{A',b',c'}=\exp \kakko{i\text{Im}\naiseki{b}{Ab'}}U_{AA',b+Ab',cc'},\]
\textit{the operators of type (S)\ }\textit{ constitute a topological group $\mathcal{G}_H$\ (with strong operator topology)\ which is isomophic to }$\mathcal{U}(H)\times H\times \mathbb{T}$\textit{ as a topological group when the latter is equipped with products }
\[(A,b,c)(A',b',c'):=(AA',b+Ab',cc'\exp \kakko{i\text{Im}\naiseki{b}{Ab'}}).\]
\textit{(Here, we topologize }$\mathcal{U}(H)$\textit{ with the strong operator topology. )}
\end{theorem}
\pf \ See \cite{Gui}.\ $\blacksquare$\\
Furthermore, we can show that type (S)\ operators are abundant.
\begin{theorem}\cite{Ism2}
\textit{The *-algebra generated by type (S) operators is weakly (strongly) dense, and hence irreducible :\ }$\{U_{A,b,c}\}''=\mathbb{B}(\Gamma(H)).$\\
\textit{More precisely,\ for any }$\varepsilon >0,$\ \textit{we have }$\{U_{I,b,1};b\in H,||b||<\varepsilon \}''=\mathbb{B}(\Gamma (H)).$
\end{theorem} 
\pf \ See \cite{Ism2}.\ $\blacksquare$\\
Therefore type (S) operators play important roles in the study of representations defined on the Fock space.
\subsubsection{Type (S) representation}
Let $V$ be a unitary representation of a topological group $\mathcal{G}$ on a Hilbert space $H$.\ A map $\beta :\mathcal{G} \to H$ is said to be a 1-cocycle of $\mathcal{G}$ w.r.t. the representation $V$\ (denoted by $\beta \in Z^1(\mathcal{G},V)$), if it satisfies $\beta (\gamma_1\gamma_2)=\beta (\gamma_1)+V(\gamma_1)\beta (\gamma_2)(\gamma_1,\gamma_2\in \mathcal{G} )$.\ 
Let $c$ be a function $c:\mathcal{G} \to \mathbb{T}$ satisfying $c(\gamma_1\gamma_2)=c(\gamma_1)c(\gamma_2)\exp \kakko{i\text{Im}\naiseki{\beta (\gamma_1)}{
V(\gamma_1)\beta (\gamma_2)}}$.\ Once $V,\ \beta$\ and $c$ are given, we can construct a unitary representation $\text{Exp }_{\beta,c}V$\ of $\mathcal{G}$ on the Boson Fock space $\Gamma(H)$\ in terms of operators of type (S).\ Such a scheme as this was proposed by Araki\ \cite{Araki2}.\ $\text{Exp }_{\beta,c}V$\ is defined as follows:
$\text{Exp}_{\beta,c}V(\gamma):=U_{V(\gamma),\beta (\gamma),c(\gamma)}$, i.e.,
\[\text{Exp }_{\beta,c}V(\gamma )\text{Exp }x=c(\gamma )\exp \kakko{-\frac{1}{2}||\beta (\gamma)||^2-\naiseki{V(\gamma)x}{\beta (\gamma)}}
\text{Exp }(V(\gamma)x+\beta (\gamma )).\]
Now we consider the special case of this construction.\ Suppose that a complex Hilbert space $H$ is the complexification of some \textbf{real} Hilbert space $H_0:\ H=H_0\otimes_{\mathbb{R}}\mathbb{C}$. Let $V_0$ be an orthogonal representation of $\mathcal{G}$ in $H_0$.\ 
Let $\beta $ be an $H_0$ valued 1-cocycle for $V$. Then we can extend $V_0$ to be a unitary representation $V$\ on the complexified Hilbert space $H$ and, in this case, $c$\ can be chosen to be a constant function 1\footnote{$\text{Im}\naiseki{V(\psi_1)\beta{(\psi_2)}}{\beta(\psi_1)}=0.$}.\ Then we obtain a unitary representation $\text{Exp}_{V,\beta}.$\ Later we will take $\mathcal{G}$\ to be the group of gauge transformations:\ $\mathcal{G}=C_c^{\infty}(M,G)=\{\psi :M\stackrel{C^{\infty}}{\to} G;\text{supp}(\psi )\text{ is compact}\ \}$\footnote{The group structure is defined by pointwise multiplications.}.\\
For $v\in H$, define a 1-coboundary $\partial v:\mathcal{G} \to H$ by $\partial v(\gamma ):=V(\gamma )v-v.$\ The set $B^1(\mathcal{G},V)$ of all 1-coboundaries for $V$ is an additive subgroup of the set of all 1-cocycles $Z^1(\mathcal{G},V)$.\ The quotient group $H^1(\mathcal{G},V)=Z^1(\mathcal{G},V)/B^1(\mathcal{G},V)$ is called a 1-cohomology group. (For more informations about this subject, see e.g. \cite{Gui, Gui2, Delorme}.)\ The unitary representations $U_{V,\beta_i}\ (i=1,2)$ constructed above are unitarily equivalent if $\beta_1$ and $\beta_2$ belong to the same cohomology class.\ That is,
if $\beta_1$ and $\beta_2$ are related by $\beta_2(\gamma)=\beta_1(\gamma)+V(\gamma)v-v$, it holds that
\[U_{I,-v,1}U_{V(\gamma),\beta_1(\gamma),1}=U_{V(\gamma),\beta_2(\gamma),1}U_{I,-v,1},\]
and the shift operator $U_{I,-v,1}$ becomes an intertwiner.\ In particular if we take a coboundary $\beta =\partial v$, then the representation $\text{Exp}_{V,\beta}$ is equivalent to $\text{Exp}_{0}V$ and the latter is easily seen to be highly reducible. (For example, the subspace $\mathbb{C}\EX{0}$ is invariant).\ Therefore in order to construct an irreducible representation we must choose a non-trivial cocycle, while the non-triviality of a cocycle does not guarantee the irreduciblity.  
\section{The energy representation of $C_c^{\infty}(M,G)$}
In this section, we review the definition of the energy representations. The gauge transformation group is defined by $C_c^{\infty}(M,G)$.\ This is considered as a group of compactly-supported sections of (trivial) fiber bundle  $P\times_{\text{Ad}}G,\ P=M\times G.$\ In $P\times_{\text{Ad}}G,$ every point is represented as $[(x,g,h)]\ (x\in M,g,h\in G)$ w.r.t.\ the equivalence relation $(x,g,h)\sim (x,g\cdot a,a^{-1}ha).$\ This group is considered as a nuclear Lie group.\footnote{For the topological properties of it, see \cite{Alb}.} 
\subsection{Isomorphism between Boson Fock space and $L^2(E',\mu)$}
There is another important realization of type (S) representations.\ Let $E$ be a real nuclear LF space i.e., a space having the topology of the inductive limit of Fr\'echet spaces. Suppose\ $E$ \ has a positive definite inner product $Q$. By the Bochner-Minlos's theorem\cite{GelV}, there is a Gaussian measure $\mu$ on the dual space $E'$ whose Fourier transform coincides with the characteristic function $\exp \kakko{-\frac{1}{2}Q(\cdot ,\cdot)}$:
\[\int_{E'}e^{i\naiseki{\chi}{F}}d\mu (\chi )=\exp \kakko{-\frac{1}{2}Q(F,F)}.\]
Let $H_0$ be a completion of $E$ w.r.t.\ the inner product $Q$, with $H:=H_0\otimes_{\mathbb{R}}\mathbb{C}$ its complexification.  
\ Then there exists a canonical isometric isomorphism $\theta $\ between the Boson Fock space $\Gamma(H)$ and the space $L^2(E',\mu;\mathbb{C})$ of complex valued square integrable functionals on $E'$\ w.r.t.\ the Gaussian measure $\mu$.\ More precisely, $\theta $ is determined by the following relation:
\[\theta \text{Exp }x=e^{\frac{1}{2}||x||^2+i\naiseki{\cdot}{x}}.\]
If $V$ is a strongly continuous orthogonal representation of a topological group $\Gamma$ on $E$ w.r.t. the inner product $Q$, $V$\ can be extended to an orthogonal representation on $H_0$. Through the complexification, it becomes a unitary representaion on $H$.\ Furthermore, it is extended to a representation on $E'$\ by the transposed action:
\[\naiseki{V(\gamma)\chi}{F}:=\naiseki{\chi}{V(\gamma)^{-1}F},\ F\in E,\ \chi \in E'.\]
Taking these facts into consideration we can transform, by the isomorphism $\theta$, the representation $\text{Exp}_{\beta}V$ into the equivalent unitary representation on $L^2(E',\mu)$.\ The transformed representation, also denoted by $\text{Exp}_{\beta}V$, is defined by
\[[\text{Exp }_{\beta}V(\gamma)\Phi ](\chi)=e^{i\naiseki{\chi}{b(\gamma)}}\Phi (V(\gamma)^{-1}\chi).\]
Historically most of the researches of the gauge group representation were based on the study of this $L^2$-space realization.
\subsection{Definition of the representation of $\MG$}
Let $M$ be a Riemannian manifold with a Riemannian metric $g$, and a Riemannian measure $dv$. Let $G$ be a compact, semisimple Lie group with Lie algebra $\mathfrak{g}$.\ $C_c^{\infty}(M,G)$\ denotes the set of $C^{\infty}$-functions from $M$ to $G$ with compact supports,\ and\ $\Omega_c^1(M,\mathfrak{g})$ the set of $\mathfrak{g}$-valued 1-forms on $M$\ with compact supports.\footnote{
Here, $\supp{\psi}:=\overline{\{x\in M;\psi(x)\neq e\}}$\ for $\psi \in \MG$\ and $\supp{\omega}:=\overline{\{x\in M;\omega_x\neq 0\}}$\ for $\omega \in \Omega_c^1(M,\mathfrak{g}).$}\ Since $\mathfrak{g}$ is semisimple, $\mathfrak{g}$ is equipped with an Ad$G$-invariant inner product defined by the minus sign of the Killing form $B(\cdot,\cdot )=\text{Tr}(\text{ad}(\cdot)\text{ad}(\cdot))$, which is negative definite by compactness.\ Next, define the inner product in $\Omega_c^1(M,\mathfrak{g})$\ as follows.\ Regard $\omega \in \Omega_c^1(M,\mathfrak{g})$ as the mapping $T(M)\to \mathfrak{g}$\ and for $x\in M$,\ define $\omega (x)^{*}$\ to be the adjoint of $\omega (x):T_x(M)\to \mathfrak{g}$ w.r.t. inner products in $T_x(M)$\ and $\mathfrak{g}$.\ Then define\footnote{We define the inner product $\naiseki{\cdot}{\cdot}$ to be anti-linear in the left variable. cf. footnotes in p. 3.} 
\[\naiseki{\omega_1}{\omega_2}:=\int_M \text{tr}\kakko{\omega_2^{*}(x)\omega_1(x)}dv(x)\]
with $\text{tr}$ a trace operator in $T_x(M)$.\\
\ \ Let $V$ be an orthogonal representation of the group $\MG$\ on the nuclear LF space $\Omega_c^1(M,\mathfrak{g})$, defined by 
\[V(\psi )\omega (x):=(\text{Ad}_{\psi (x)})_{*}(\omega (x)),\ \ \ (x\in M,\omega \in \Omega_c^1(M,\mathfrak{g}),\psi \in \MG ).\]
A distinguished 1-cocycle of $\beta:\ \MG \to \Omega_c^1(M,\mathfrak{g})$\ of $V$, called  the Maurer-Cartan cocycle is defined by 
\[\beta (\psi )(x):=d\psi (x)\cdot \psi (x)^{-1}\kakko{=(R_{\psi(x)^{-1}})_{*\ \psi (x)}(d\psi _x(\cdot ))}:T_x(M)\to \mathfrak{g}.\]
Let $H_0$ be the completion of $\Omega_c^1(M,\mathfrak{g})$ w.r.t. $\naiseki{}{}$,\ $H:=H_0\otimes_{\mathbb{R}}\mathbb{C}$\ its complexification.\\
From the previous argument, we obtain a unitary representation $U=\text{Exp}_{V,\psi}$\ of $\MG$\ on the Boson Fock space $\Gamma(H)$:
\[U(\psi)\text{Exp }\omega :=e^{-\frac{1}{2}||\beta(\psi)||^2-\naiseki{V(\psi)\omega}{\beta (\psi)}}
\text{Exp}\kakko{V(\psi )\omega +\beta (\psi)}.\]
To conclude this section, we state some properties of 1-cocycles for the representation $V$.
\begin{proposition}
\textit{Let $P_{\psi}^V$ be an orthogonal projection onto the subspace $\{\omega \in H;\ V(\psi)\omega =\omega\}$.\ Let \ $\psi,\psi_1,\psi_2 ,\psi_3\ \in \MG$.\ It holds for $\gamma \in Z^1(\MG,V)$ that}\\
(1)\ \textit{$\supp{\gamma(\psi)}\subset \supp{\psi}.$}\\
(2)\ If $\psi_1=\psi_2$\ on an open subset $U\subset M$, $\gamma(\psi_1)=\gamma(\psi_2)$\ on $U$.\\
(3)\ \textit{$\dis \limn \frac{1}{n}\gamma(\psi_1\psi_2^{n}\psi_3)=V(\psi_1)P_{\psi_2}^V\gamma(\psi_2).$}\\
(4)\ \textit{If $\dis \limn \frac{1}{n}\gamma (\psi^n)\neq 0$ for some $\psi$\footnote{This limit exists. Put $\psi_1=\psi_3=1,\ \psi_2=\psi.$\ in (3).}, then $\gamma $ is not a trivial cocycle.\ In particular, Maurer-Cartan cocycle\ $\beta$\ is not trivial:\ $\beta \notin B^1(H,V).\ (\textit{In fact}\ \beta \notin \overline{B^1(H,V)}$.)} 
\end{proposition}
Note that the limit in (3) does not depend on $\psi_3$.\\
\pf \ (1)\ Let $K:=\supp{\psi}.$\ and $x\notin K.$\ We shall prove $\gamma(\psi)(x)=0.$\ Since $K$ is closed, there is some compact $K_1$ such that $x\in K_1,\ K_1\cap K=\phi.$ Let $\psi_1\in \MG$ be any function whose support is contained in $K_1$. Since $\psi$\ and $\psi_1$ have disjoint supports, we have $\psi\psi_1=\psi_1\psi$.\ Then from the 1-cocycle condition, it holds that
\[\gamma(\psi\psi_1)=\gamma(\psi_1\psi),\]
or equivalently
\[\gamma(\psi)+V(\psi)\gamma(\psi_1)=\gamma(\psi_1)+V(\psi_1)\gamma(\psi).\]
Since $\supp{\psi}\cap K_1=\phi, V(\psi)=I$ on $K_1.$\ Therefore on $K_1$, we have
\[\gamma(\psi)+\gamma(\psi_1)=\gamma(\psi_1)+V(\psi_1)\gamma(\psi).\]
Therefore
\[\gamma(\psi)=V(\psi_1)\gamma(\psi).\]
Suppose $\gamma(\psi)(x)\neq 0.$\ Then due to the triviality of the center of $\mathfrak{g}$, there is some $\psi_1,\ \supp{\psi_1}\subset K_1$ such that $V(\psi_1)\gamma(\psi)(x)\neq \gamma(\psi)(x),$\ which contradicts the above equality.\ Therefore $\gamma(\psi)(x)=0$\ and $\supp{\gamma(\psi)}\subset \supp{\psi}.$\\
(2)\ From the 1-cocycle condition again, we have
\eqa{
\gamma(\psi_2\psi_1^{-1})&=\gamma(\psi_2)+V(\psi_2)\gamma(\psi_1^{-1})\\
&=\gamma(\psi_2)+V(\psi_2)[-V(\psi_1^{-1})\gamma(\psi_1)]\\
&=\gamma(\psi_2)-V(\psi_2\psi_1^{-1})\gamma(\psi_1)\ \ \ \textrm{on }U.
}
Note that $\gamma(1)=0,\ $which implies $\gamma(\psi^{-1})=-V(\psi^{-1})\gamma(\psi).$\ Since $U$ is open and $\psi_1|_U=\psi_2|_U$, we have $\supp{\psi_2\psi_1^{-1}}\cap U=\phi.$ Therefore from (1), it holds that $\supp{\gamma(\psi_2\psi_1^{-1})}\cap U=\phi$. Thus, we obtain $\gamma(\psi_2\psi_1^{-1})=0$ on $U$.\ Furthermore, $V(\psi_2\psi_1^{-1})=\text{id}$ on $U$.\ Therefore we have
\eqa{
0&=\gamma(\psi_2)-V(\psi_2\psi_1^{-1})\gamma(\psi_1)\\
&=\gamma(\psi_2)-\gamma(\psi_1)\ \ \ \textrm{on }U.
}
(3)\ First, we prove it for $\psi_1=1$ case.\ From the 1-cocycle condition, we have
\eqa{
\gamma (\psi_2^{n}\psi_3)&=\gamma(\psi_2)+V(\psi_2)\gamma(\psi_2^{n-1}\psi_3)\\
&=\gamma(\psi_2)+V(\psi_2)(\gamma(\psi_2)+V(\psi_2)\gamma(\psi_2^{n-2}\psi_3))\\
&=(I+V(\psi_2))\gamma(\psi_2)+V(\psi_2)^2[\gamma(\psi_2)+V(\psi_2)\gamma(\psi_2^{n-3}\psi_3)]\\
&=\cdots \\
&=(I+V(\psi_2)+V(\psi_2)^2+\cdots +V(\psi_2)^{n-2})V(\psi_2)+V(\psi_2)^{n-1}[\gamma(\psi_2)+V(\psi_2)\gamma(\psi_3)],\\
\frac{1}{n}\gamma(\psi_2^n\psi_3)&=\frac{1}{n}\sum_{k=0}^{n-1}V(\psi_2)^k\gamma(\psi_2)+\frac{1}{n}V(\psi_2)^n\gamma(\psi_3)\\
&\stackrel{n\to \infty}{\longrightarrow }P_{\psi_2}^V\gamma(\psi_2).
}
Here in the last equality we used the von Neumann's mean ergodic theorem\ (\cite{ReedSimon}, p. 57).\ The general $\psi_1$\ case follows from the $\psi_1=1$\ case:\ 
\eqa{
\frac{1}{n}\gamma(\psi_1\psi_2^n\psi_3)&=\frac{1}{n}\{\gamma(\psi_1)+V(\psi_1)\gamma(\psi_2^n\psi_3)\}\\
&\stackrel{n\to \infty}{\longrightarrow }V(\psi_1)P_{\psi_2}^V\gamma(\psi_2).
}
(4)\ For any trivial cocycle $\partial \omega\ (\omega \in H),$ it holds that
\eqa{
\frac{1}{n}||\partial \omega(\psi^n)||&=\frac{1}{n}||(V(\psi^n)-I)\omega||\\
&\le \frac{2}{n}||\omega ||.
}
Therefore $\dis \limn \frac{1}{n}\partial \omega(\psi^n)=0$\ and the claim holds.\ Next we prove $\beta \notin B^1(H,V).$\ Fix a nonzero abelian subalgebra $\mathfrak{h}$\ and
consider $\varphi \in C_c^{\infty}(M,\mathfrak{h}),\ d\varphi \neq 0.$\ Since $\psi=e^{\varphi}$\ takes values in an abelian subgroup of $G$, we have $\beta (\psi )=d\varphi$.\ Then we get
\eqa{
\frac{1}{n}\beta(e^{n\varphi})&=\frac{1}{n}\cdot nd\varphi \\
&=d\varphi \ (\neq 0).
}
Therefore $\beta \notin B^1(H,V).\ \blacksquare$\\
We add an alternative proof of it.\\
\textbf{Second proof of $\beta \notin B^1(H,V)$.} \\
Suppose $\beta (\psi)=\partial \omega (\psi)$ for some $\omega \in H.$\ For $s\in \mathbb{R},$ we have 
\eqa{
\beta(e^{s\varphi})=sd\varphi&=(V(e^{s\varphi})-I)\omega\ (s\in \mathbb{R}-\{0\})\\
\Leftrightarrow d\varphi&=\frac{1}{s}[V(e^{s\varphi})-I]\omega \\
&\stackrel{s\to 0}{\longrightarrow }[\varphi,\omega],\ \forall \varphi \in C_c^{\infty}(M,\mathfrak{h}).
}
Here in the last equality we have used the formula 
\[\dis V(e^{s\varphi})\omega =e^{s[\varphi ,\cdot ]}\omega =\omega +s[\varphi,\omega ]+\frac{s^2}{2!}[\varphi,[\varphi,\omega ]]+\cdots .\] 
Since $\mathfrak{g}$ is semisimple, we can show that for any compact set $K \subset M$ there exists such $\varphi \in \Mg$\ as is constant on $K$ but $[\varphi,\omega]\neq 0$\ on some nonempty open subset of $K$.\ This is clearly a contradiction $.\ \blacksquare$\\
\textbf{Remark.}\\
Since there is no nonzero $\Omega \in \Gamma(H)$\ that is invariant under all $U(\psi)$'s, we can prove that 
$\supp{\gamma(\psi)}\subset \supp{\psi}$\ for $\gamma \in Z^1(\MG,\ U).$
\section{Local structures of the net $\dom \mapsto \mathcal{M}(\dom)''$}
In this section the results obtained by the present author are explained in details and are proved.\ Our discussion is based on the algebraic structure of type (S) opertors and the support properties of $\psi \in \MG$.\ The type (S) relations:
\[U_{A,b,c}U_{A',b',c'}=\exp \kakko{i\ \text{Im}\naiseki{b}{Ab'}}U_{AA',b+Ab',cc'},\]
reminds us of the commutation relations of Weyl unitaries :
\[W(h)W(k)=\exp \kakko{-\frac{i}{2}\text{Im }\naiseki{h}{k}}W(h+k).\]
(In fact the latter is a special form of the former with some modifications.)\ Therefore it is useful to compare their algebraic structures. Although our representation is not a genuine quantization of gauge fields yet, it may shed some lights on the possible structure of quantum gauge field theory. First, we briefly describe the structure of Weyl unitaries, which is a representation of CCR's of free Bose fields. 
\subsection{Local structure of Weyl unitaries}
Consider a neutral scalar field. They are generated by Weyl unitaries.\\
Let $H$\ be a \textbf{complex} Hilbert space, $K\subset H$\ a closed \textbf{real} subspace of $H$\ (for real subspace $K$, we denote by $K\le_{\mathbb{R}}H$).\ We define Weyl unitary operators $W(h)(h\in H)$\ on $\Gamma (H)$. They are determined by
\[\begin{cases}
W(h)W(k)=\exp \kakko{-\frac{i}{2}\text{Im }\naiseki{h}{k}}W(h+k)\\
W(h)\EX{0}=\exp \kakko{-\frac{1}{4}||h||^2}\EX{\frac{ih}{\sqrt{2}}}.\end{cases}\]
The von Neumann algebra $\mathcal{M}(K):=\{W(h);h\in K\}''$\ is called a second quantization algebra. Here, $\mathcal{S}'=\{T\in \mathbb{B}(\Gamma (H));TS=ST, \forall S\in \mathcal{S}\}$\ is the commutant of $\mathcal{S}\subset \mathbb{B}(\Gamma (H))$.\ 
For $\mathcal{K}\subset H,$\ define $\mathcal{K}':=\{h\in H;\text{Im}\naiseki{h}{k}=0,\forall k\in K\}$\ (symplectic complement of $M$). For a general von Neumann algebra $\mathcal{M}$ acting in a Hilbert space $H$, we say a vector $\Omega \in H$ is cyclic for $\mathcal{M}$ in $H$\ if $\mathcal{M}\Omega $ is dense in $H$. We also say $\Omega $\ is separating for $\mathcal{M}$ in $H$ if for any $Q\in \mathcal{M}$, $Q\Omega =0\Leftrightarrow Q=0$ holds. This condition is equivalent to the cyclicity of $\Omega$ for $\mathcal{M}'$ in $H$.\ 

The following properties hold.
\begin{theorem}\cite{Araki1,Guido,RobLeyTes} \textit{For} $K\le_{\mathbb{R}}H,$\ \textit{we have}\\
(1)\ $K'$\ \textit{is a closed real subspace of }$H$.\\
(2)\ $K_1\subset K_2\Rightarrow K_1'\supset K_2'$\\
(3)\ $K''$\ \textit{is the closed real subspace of }$H$\ \textit{generated by }$K$.\\
(4)\ $(K+iK)'=K'\cap iK'$= \textit{the set of all vectors orthogonal to }$K$.\\
(5)\ $K'=\{0\}$ if $K$\ \textit{is a dense subspace of }$H$.\\
(6)\ \textit{For} \textit{a closed real subspace}\ $K$\ \textit{of }$H$\ \textit{ and an orthogonal projection}\ $P$\ \textit{in}\ $H$, \textit{the following quivalence holds:}
\[PK\subset K\Leftrightarrow (I-P)K\subset K\Leftrightarrow PK'\subset K'\Leftrightarrow (I-P)K'\subset K'.\]
\textit{If one of these conditions is valid, then} 
\[P(K')=(PK)'\cap PH.\]
\end{theorem}
Second quantization algebras has a natural modular structure.\ The subspace $K\le_{\mathbb{R}}H$\ is called standard if 
$K+iK\text{ is dense in }H$ and $K\cap iK=\{0\}.$\ If $K$\ is standard, then we can define the canonical involution $s:K+iK\to K+iK$\ by $s:h+ik\mapsto h-ik,\ h,\ k\in K$.\ It can be shown that
\begin{theorem}\cite{RobLeyTes} \textit{If $K$\ is standard in $H$, then }\\
(1)\ \textit{$s$\ is a densely defined,\ closed antilinear involution.}\\
(2)\ \textit{$K'$\ is also standard and the canonical involution is the adjoint $s^*$\ of $s$.}\\
(3)\ \textit{If $s=j\delta^{\frac{1}{2}}$\ is the polar decomposition of $s$,\ then}
\[j^2=I,j\delta^{\frac{1}{2}}=\delta^{-\frac{1}{2}}j,j(K)=K'.\]
\end{theorem}
Before the birth of Tomita-Takesaki theory, Araki\ \cite{Araki1} showed that\ (when stated in the modern style)
\begin{theorem}\cite{Araki1,RobLeyTes} \textit{The vacuum vector}\ $\EX{0}$\ \textit{is cyclic and separating for }$\mathcal{M}(K)\ [\mathcal{M}(K)$ \textit{is in a standard form w.r.t. the Fock vacuum }]\ \textit{iff}
$K$\ \textit{is standard.\ More precisely, the following statements hold for}\ $K\le_{\mathbb{R}}H$.\\
(1)\ $\mathcal{M}(K)=\mathcal{M}(\overline{K}).$\\
(2)\ $\EX{0}$\ \textit{is cyclic for }$\mathcal{M}(K)$\ \textit{iff }$K+iK$\ \textit{is dense in }$H$.\\
(3)\ $\EX{0}$\ \textit{is separating for }$\mathcal{M}(K)$\ \textit{iff }$\overline{K}\cap i\overline{K}=\{0\}.$\\
(4)\ $\mathcal{M}(K)'=\mathcal{M}(K')$\ \textit{(Haag duality)}.\\
(5)\ $\{\mathcal{M}(K_1)\cup \mathcal{M}(K_2)\}''=\mathcal{M}(K_1+K_2)$.\\
(6)\ $\mathcal{M}(K_1)\cap \mathcal{M}(K_2)=\mathcal{M}(\overline{K_1}\cap \overline{K_2})$.\ \textit{and consequently }$\mathcal{M}(K)$\ \textit{is a factor iff }$\overline{K}\cap iK'=\{0\}.$
\end{theorem}  
From the above theorem, we can define the densely defined operator in $H$ with a cyclic and separating vector $\Omega:=\EX{0}$.
\[S_0:\mathcal{M}(K)\Omega \to \mathcal{M}(K)\Omega ,A\Omega \mapsto A^*\Omega,\]
which is known to be closable.\ Furthermore, if we consider the polar decomposition of the closure $S$\ of $S_0,$
\[S=J\Delta^{\frac{1}{2}},\]
we then arrive at the following theorem of Osterwalder-Eckman\cite{Osterwalder}, which is a reformulation of Araki's result in the language of Tomita-Takesaki modular theory.
\begin{theorem}\cite{Osterwalder, RobLeyTes}\ $S=e^s,J=e^j,\Delta =e^{\delta}.$\ \textit{Here, $e^A$\ is a second quantization of a (possibly unbounded) operator $A\ in\ H.$}
\end{theorem}
In summary, there is a natural modular-symplectic structure in the algebra of Weyl unitaries. In particular, the Fock vacuum is cyclic and separating for any proper local subalgebra. In fact the local algebra is proven to be a unique injective type III$_1$ factors\ (cf.\ \cite{Haag,Guido}). In the next subsection we compare these results with the corresponding local gauge algebras.
\subsection{Local structure of the representation of gauge group}
Next, we state the corresponding local structures of gauge group representation. To this end, let us introduce some notations.\ Let $U_{A,b}:=U_{A,b,1}$ and $\mathcal{A}=\text{Lin}\{U_{A,b};(A,b)\in \mathcal{U}(H)\times H\}.$\ Let $\mathcal{O}$\ be an open subset of $M$.\ We consider the net $\mathcal{O}\mapsto \mathcal{M}(\mathcal{O})''$\ of von Neumann algebras generated by the *-algebras\ $\mathcal{M}(\mathcal{O}):=\text{Lin}\{U(\psi);\text{supp}(\psi)\subset \mathcal{O}\}.$\ We also consider the *-algebra defined by 
\[\mathcal{A}(\mathcal{O}):=\text{Lin}\{U_{A,b};A|_{H(\mathcal{O'})}=\text{Id}_{H(\mathcal{O'})},\ AH(\dom)\subset H(\dom),\ \text{Int}(\text{supp}(b)\cap \mathcal{O'})=\phi\ \}.\]
Here, $\dom':=M\backslash \dom$\ is the complement of $\dom$.\ Now we state our main results.
\begin{theorem}\label{main}
\textit{ Let $N_0$ be a neighborhood of $1\in \MG$.\ For an open subset \ $\mathcal{O}\subset M,$\ we have}\\
(1)\ \textit{$\mathcal{M}(\mathcal{O})'\cap \mathcal{A}=\mathcal{A}(\mathcal{O'}).$\ In particular,
$\mathcal{M}'\cap \mathcal{A}=\mathbb{C}1.\ \mathcal{M}:=\mathcal{M}(M).$\ Furthermore, it holds that $U(N_0)'\cap \mathcal{A}=\mathbb{C}1$.}\\
(2)\ \textit{The net $\mathcal{O}\mapsto \mathcal{M}(\mathcal{O})''$\ satisfies} 
\[\begin{cases}
isotony:\ \mathcal{O}_1\subset \mathcal{O}_2\Rightarrow \mathcal{M}(\mathcal{O}_1)''\subset \mathcal{M}(\mathcal{O}_2)''.\\
locality:\ \mathcal{O}_1\subset \mathcal{O}_2'\Rightarrow \mathcal{M}(\mathcal{O}_1)''\subset \mathcal{M}(\mathcal{O}_2)'.\\
\dis additivity:\ M=\UU_{i}\mathcal{O}_i\Rightarrow \mathcal{M}''=\kakko{\UU_i \mathcal{M}(\mathcal{O}_i)}''.
\end{cases}\]
\textit{However, the Fock vacuum $\Omega :=\EX{0}$\ is not cyclic for $\mathcal{M}(\mathcal{O})''$if $\dom \neq \phi, M$.}\\
\end{theorem}
To prove this theorem, we need some lemmata.
\begin{lemma}\label{lem1} \textit{If}$\ \{(A_i,b_i)\}_{i=1}^N(N\geqq 2)$ \textit{are elements of $\mathcal{U}(H)\times H$,\ any two of which are different, then there exists a number $i_0(1\le i_0\le N)$ and a vector $x_{i_0}\in H$ such that $A_{i_0}x_{i_0}+b_{i_0}\neq A_jx_{i_0}+b_j,(\forall j\neq i_0).$}
\end{lemma}
\pf \ We prove this lemma by induction. It is obvious for $N=2$.\ Suppose we have proven for $N\ (N\ge 2)$.\ Let us consider the $(N+1)$ elements $\{(A_i,b_i)\}_{i=1}^{N+1}$. Then there exists some $(i_0,x_{i_0})$ such that  $A_{i_0}x_{i_0}+b_{i_0}\neq A_jx_{i_0}+b_j,(\forall j\neq i_0,1\le j\le N).$\ Compare $(A_{N+1},b_{N+1})$ and $(A_{i_0},b_{i_0})$.\\
(i)\ $b_{i_0}=b_{N+1}$ case.\\
If $A_{N+1}x_{i_0}\neq A_{i_0}x_{i_0}$, then $(i_0,x_{i_0})$ satisfies the requirement.\ If not, since $A_{N+1}\neq A_{i_0}$, there exists $y\neq 0\in H$ such that $A_{N+1}y\neq A_{i_0}y.$\\
As $A_{i_0}x_{i_0}+b_{i_0}\neq A_jx_{i_0}+b_j$, we can take\ $\varepsilon >0$\ so small that we obtain
\[A_{i_0}(x_{i_0}+\varepsilon y)+b_{i_0}\neq A_j(x_{i_0}+\varepsilon y)+b_j, (j\neq i_0,j\leqq N).\]
And this is also valid for $j=N+1.$\\
(ii)\ $b_{i_0}\neq b_{N+1}$ case.\\
If $A_{N+1}x_{i_0}+b_{N+1}\neq A_{i_0}x_{i_0}+b_{i_0}$, then $(i_0,x_0)$ satisfies the requirement.\\
If $A_{N+1}x_{i_0}\neq A_{i_0}x_{i_0}$ and furthermore, $A_{N+1}x_{i_0}+b_{N+1}=A_{i_0}x_{i_0}+b_{i_0}$, then for $\varepsilon >0$, $A_{N+1}(1+\varepsilon )x_{i_0}+b_{N+1}\neq A_{i_0}(1+\varepsilon )x_{i_0}+b_{i_0}$ and we can take $\varepsilon$
so small that 
$A_{i_0}(1+\varepsilon )x_{i_0}+b_{i_0}\neq A_j(1+\varepsilon )x_{i_0}+b_j$ still holds $.\ \blacksquare$
\begin{lemma}\label{indep of (S)}\textit{Type (S) operators $\{U_{A,b}\}$ are linearly independent.}
\end{lemma}
\pf \ Suppose $\dis \sum_{i=1}^N\lambda_iU_{A_i,b_i}=0$. Again we prove this by induction on $N$.\ We may assume $(A_i,b_i)\neq (A_j,b_j)\ (i\neq j).$\ For any $x\in H$, it follows that
\[\sum_{i=1}^N\lambda_ie^{-\frac{1}{2}||b_i||^2-\naiseki{A_nx}{b_n}}\text {Exp}(A_nx+b_n)=0\]
(i)\ $N=2$ case.\\
Since exponential vectors are linearly independent and $(A_1,b_1)\neq (A_2,b_2)$, there exists some $x\in H$ such that $A_1x+b_1\neq A_2x+b_2$. Therefore $\text{Exp }(A_ix+b_i)$ are linearly independent and we obtain $\lambda_1=\lambda_2=0.$\\
(ii)\ Suppose we have proven for $N-1$.\ Then by the last lemma there exists some $x_{i_0}$ such that $A_{i_0}x_{i_0}+b_{i_0}$\ is different from any of $A_jx_{i_0}+b_j(j\neq i_0)$.\ Therefore again by independence we obtain $\lambda_{i_0}=0$.\ Thus from the induction hypotheses all of the $\lambda_i$ are 0\ $.\ \blacksquare$\\
\begin{lemma}\cite{Gel77,Alb81,Wal}\label{totality}
\textit{$\{V(\psi)d\varphi ;\text{supp}(\psi),\text{supp}(\varphi)\subset \mathcal{O}\}$\ is a total set in $H(\mathcal{O}).$}
\end{lemma}
\pf \ This is a slight modification of the fact proved in \cite{Alb81}.\ $\blacksquare$\\
\begin{lemma}\label{beta}\ \textit{ $\{\beta(\psi);\text{supp}(\psi)\subset \mathcal{O}\}$\ is also total in $H(\mathcal{O})$.}
\end{lemma}
\pf \ This is a consequence of the 1-cocycle property of $\beta$.\ Let $\mathfrak{h}$\ be any Cartan subalgebra of $\mathfrak{g}.$\\
For $\varphi \in C_c^{\infty}(M,\mathfrak{h}),\ \psi \in C_c^{\infty}(M,G)$\ with $\text{supp}(\psi),\ \text{supp}(\varphi)\subset \mathcal{O},$\ we have
\[\beta(\psi e^{\varphi})=\beta (\psi)+V(\psi)d\varphi.\]
Therefore
\[V(\psi)d\varphi =\beta (\psi e^{\varphi})-\beta (\psi)\in \text{Lin}\{\beta (\psi);\text{supp}(\psi)\subset \mathcal{O}\}\tag{$\diamondsuit$}\]
From the totality of Cartan subalgebras as discussed above, we see that $(\diamondsuit )$\ holds for any $\varphi \in C_c^{\infty}(M,\mathfrak{g}),\text{supp}(\varphi)\subset \mathcal{O}.$\ Thus, the claim holds from the totality of $\{V(\psi)d\varphi \}\ .\ \blacksquare$\\
\textbf{Remark.}\\
Note that for $\varphi \in C_c^{\infty}(M,\mathfrak{h}),$
\[\beta (e^{\varphi})=d\varphi=\textbf{exact 1-form.}\]
However, the proposition says that even if $M$ has non-trivial de Rham cohomology\ (with compact support)\ $H_c^1(M,\mathbb{R})\neq 0$, $\{\beta (\psi )\}$\ is a total set in $H$.\ This is a consequense of the Lie algebra structure. Therefore although abelian one-parameter subgroups $s\mapsto e^{s\varphi},\ \varphi \in C_c^{\infty}(M,\mathfrak{h})$\ play important roles in our analysis, they are not sufficient for understanding the whole structure of the Maure-Cartan cocyle.
\begin{lemma}\label{separation}\ \textit{Let $\psi \in \MG,\ \supp{\psi}\subset \mathcal{O},\ \varphi \in \Mg,\ \supp{\varphi}\subset \mathcal{O},\ K:=\supp{d\varphi}.$ Let $N_0(\mathcal{O})$ be a neighborhood of $1\in \MG$ whose elements are supported in $\dom$.\ Then there exist an open covering $\{V_k\}_{k=1}^N$\ of $K$\ and two families of functions $\{\psi_j^k\}_{0\le j\le n_k<\infty, 1\le k\le N}\subset N_0(\mathcal{O}),\ \{\varphi_k\}_{1\le k\le N}\subset \Mg,\ \supp{\varphi_k}\subset V_k$\ such that}
\[
\begin{cases}
\dis \psi|_{V_k}&=\psi_{n_k}^k\psi_{n_k-1}^k\cdots \psi_{0}^k|_{V_k},\ \forall k,\\
d\varphi &= \dis \sum_{k=1}^Nd\varphi_k\ \text{on}\ K.
\end{cases}\]
\end{lemma}
\pf \ Let $x\in K,\ g:=\psi (x)$.\ There exist an open neighborhood $U_x$ of $x$\ and some function $\psi_0^x\in N_0(\mathcal{O})$\ such that $\psi(y)=g\cdot \psi_0^x(y),\forall y\in U_x.$\ It is known that for any connected neighborhood $\mathcal{E}$\ of $e\in G$, we have $G=\UU_{k\ge 1}\mathcal{E}^k,$\ where $\mathcal{E}^k:=\{g_1g_2\cdots g_k;\ g_i\in \mathcal{E},\ \forall i\}$.\ Taking a small $\mathcal{E},$\ there exists $\{g_i\}_{i=1}^{l}\subset \mathcal{E},\ \{\psi_i^x\}_{i=1}^{l}\subset N_0(\mathcal{O})$\ and neighborhoods $V_{i,x}$\ of $x,V_{i,x}\subset U_x$,\ such that $g=g_l\cdot g_{l-1}\cdots g_1$\ and $\psi_i^x(y)=g_i,\ \forall y\in V_{i,x}$.\ Define $V_x:= \cap_iV_{i,x}\neq \phi.$\ We see that
\[\psi|_{V_x}=\psi_{l}^x\cdots \psi_0^x|_{V_x}.\] 
Since $K\subset \UU_{x\in K}V_x$\ and $K$\ is compact, there exists $x_1,\cdots ,x_N\in K$\ such that $K\subset \UU_{k=1}^NV_{x_k}$.\ Define $V_k:=V_{x_k}.$\ Let $\{\chi_k\}_{k=1}^N$\ be a partition of unity associated with the covering $\{V_k\}$\ of $\UU_kV_k$.\ Take $\varphi_k:=\chi_k\varphi$\ and we obtain the result$.\ \blacksquare$\\
\textbf{Proof of Theorem\ref{main}}\\
(1)\ 
We can write $\Xi \in \mathcal{M}(\mathcal{O})'\cap \mathcal{A}$ as $\Xi=\sum_{i=1}^N\lambda_iU_{A_i,b_i}$, where all of $(A_i,b_i)$'s are different and $\lambda_i\neq 0,\forall i.$\ Since $\Xi \in \mathcal{M}(\mathcal{O})'$, we have $U(\psi)\Xi U(\psi)^{-1}=\Xi$\ for $\text{supp}(\psi)\subset \mathcal{O}.$\ Therefore  
\eqa{
\sum_{i}\lambda_iU(\psi)U_{A_i,b_i}U(\psi)^{-1}&=\sum_{i}\lambda_iU_{V(\psi),\beta (\psi)}U_{A_i,b_i}U_{V(\psi)^{-1},-V(\psi)^{-1}\beta (\psi)}\\
&=\sum_ie^{i\text{Im}\naiseki{\beta(\psi)}{V(\psi)b_i}}\lambda_iU_{V(\psi)A_i,\beta (\psi)+V(\psi)b_i}U_{V(\psi)^{-1},-V(\psi)^{-1}\beta (\psi)}\\
&=\sum_ie^{i\theta(\psi,A_i,b_i)}\lambda_iU_{V(\psi)A_iV(\psi)^{-1},\beta(\psi)+V(\psi)b_i-V(\psi)A_iV(\psi)^{-1}\beta(\psi)}\\
&=\sum_i\lambda_iU_{A_i,b_i},
}
where\ $\theta(\psi,A_i,b_i):=\text{Im}\{\naiseki{\beta(\psi)}{V(\psi)b_i}-\naiseki{\beta(\psi)+V(\psi)b_i}{V(\psi)A_iV(\psi)^{-1}\beta(\psi)}\}.$\\
(Note that $U_{A,b}U_{A',b'}=e^{i\text{Im}\naiseki{b}{Ab'}}U_{AA',b+Ab'},\ U_{A,b}^{-1}=U_{A^{-1},-A^{-1}b}.$)\\
From the Hausdorff property there exist small neighborhoods $W_i$ of $(A_i,b_i)$ such that $W_i\cap W_j=\phi (i\neq j)$. From the continuity of $\psi \mapsto U(\psi)$, there exists some small neighborhood $N_0$ of a constant function $1\in \ C_c^{\infty}(M,G)$ such that for any $\psi \in N_0(\mathcal{O}):=\{\psi \in N_0;\text{supp}(\psi)\subset \mathcal{O}\}$,\ $(A_i^{\psi},b_i^{\psi})\in W_i$\ holds. Here, we define 
\[\begin{cases}
A_i^{\psi} &:=V(\psi)A_iV(\psi)^{-1},\\
b_i^{\psi} &:=\beta(\psi)+V(\psi)b_i-V(\psi)A_iV(\psi)^{-1}\beta(\psi). 
\end{cases}\]
Suppose for some $\psi \in N_0$ and $i_0$,\ $(A_{i_0}^{\psi},b_{i_0}^{\psi})\neq (A_{i_0},b_{i_0})$. From Lemma\ \ref{indep of (S)}, $\{U_{A,b}\}$'s are independent.\ Hence we obtain $\lambda_{i_0}=0$, which is a contradiction. Therefore for all $i$ and $\psi \in N_0(\mathcal{O}),(A_i^{\psi},b_i^{\psi})=(A_i,b_i)$. Furthermore, we have $\theta(\psi,A_i,b_i)\in 2\pi \mathbb{Z}.$\ Now, from the fact that $\beta(1)=0,$  we see that $\theta(1,A_i,b_i)=\text{Im}\{\naiseki{0}{b_i}-\naiseki{0+b_i}{0}\}=0.$
\ Since $\psi \mapsto \theta (\psi,A_i,b_i)$\ is continuous, it holds that $\theta (\psi,A_i,b_i)= 0.$\ Therefore the proof is reduced to the following proposition. 
\begin{proposition}
$A_i|_{H(\mathcal{O})}=\text{Id}_{H(\mathcal{O})},\ \text{Int}(\text{supp}(b)\cap \mathcal{O})=\phi (\forall i),\ A_iH(\dom')\subset H(\dom').$
\end{proposition}
\pf \ From the above argument, it follows that for all $\psi \in N_0(\mathcal{O}),$
\eqa{
V(\psi)A_iV(\psi)^{-1} &= A_i,\ \tag{$\flat$}\\
\beta(\psi)+V(\psi)b_i-V(\psi)A_iV(\psi)^{-1}\beta(\psi)  &= b_i. \tag{$\natural$}
}
Insert $(\flat )$ into $(\natural )$, then we obtain 
\[\beta (\psi )+V(\psi )b_i-A_i\beta (\psi )=b_i.\ \tag{$\sharp$}\]
Next, let us take an arbitrary Cartan subalgebra $\mathfrak{h}$ of a semisimple Lie algebra $\mathfrak{g}$ and consider the map $\varphi \in C_c^{\infty}(M,\mathfrak{h}),\ \text{supp}(\varphi )\subset \mathcal{O}$,  and define $\psi :=e^{s\varphi }(s\in \mathbb{R})$.\ For sufficiently small $|s|$, $\psi $ belongs to $N_0(\mathcal{O})$.\ Since $\mathfrak{h}$ is commutative, we obtain $\beta (\psi)=sd\varphi$.\ Therefore in this case $(\sharp )$ is reduced to
\eqa{
sd\varphi +V(e^{s\varphi })b_i-sA_id\varphi = b_i&\Leftrightarrow (A_i-I)d\varphi =\frac{V(e^{s\varphi})-I}{s-0}b_i.
}
In the $s\to 0$ limit, it follows that
\[(A_i-I)d\varphi =[\varphi,b_i]\ \tag{$\spadesuit$}\]
for all $\varphi \in C_c^{\infty}(M,\mathfrak{h}),\ \text{supp}(\varphi)\subset \mathcal{O}.$\ Since the whole Lie algebra $\mathfrak{g}$ is a union of all Cartan subalgebras:\ $\dis \mathfrak{g}=\UU_{\text{Cartan subalgs}}\mathfrak{h}$ and the equality\ $(\spadesuit )$ is linear in the variable $\varphi$, it is valid for \textbf{all} $\varphi \in C_c^{\infty}(M,\mathfrak{g}),\ \text{supp}(\varphi)\subset \mathcal{O}$.\\
\ Next, we prove $\text{Int}(\text{supp}(b)\cap \mathcal{O})=\phi$.\\
Consider again arbitrary Cartan subalgebra $\mathfrak{h}$\ and the corresponding root space decomposition of the complexification of semisimple $\mathfrak{g}:$
\[\mathfrak{g}_{\mathbb{C}}=\mathfrak{h}_{\mathbb{C}}\oplus \OP_{\alpha \in \Delta}\mathfrak{g}_{\alpha}.\]
Here, $\Delta$\ is the root system for $\mathfrak{h}$\ and $\mathfrak{g}_{\alpha}$\ is the root space corresponding to $\alpha \in \Delta.$\ Let $\omega_{\alpha}\in C_c^{\infty}(M,\mathfrak{g}_{\alpha}),\ \supp{\omega_{\alpha}}\subset \dom.$\ Put $\psi :=e^{\varphi}\in N_0(\mathcal{\dom}),\ \varphi \in C_c^{\infty}(M,\mathfrak{h}).$\ From $V(\psi)A_i=A_iV(\psi),$\ it follows that
\eqa{
V(e^{\varphi})A_i\omega_{\alpha}&=A_iV(e^{\varphi})\omega_{\alpha}\\
&=A_i[\omega_{\alpha}+i\naiseki{\alpha}{\varphi}\omega_{\alpha}+\frac{1}{2!}(i\naiseki{\alpha}{\varphi})^2\omega+\cdots ]\\
&=e^{i\naiseki{\alpha}{\varphi}}A_i\omega_{\alpha}.
}
This equality implies $A_i\omega_{\alpha}\in C_c^{\infty}(M,\mathfrak{g}_{\alpha}).$\ Therefore we see that $A_i$\ preserves the root space structures:
\[\begin{cases}
C_c^{\infty}(M,\mathfrak{h})\stackrel{A_i}{\longrightarrow } C_c^{\infty}(M,\mathfrak{h}),\\
C_c^{\infty}(M,\mathfrak{g}_{\alpha})\stackrel{A_i}{\longrightarrow } C_c^{\infty}(M,\mathfrak{g}_{\alpha}).
\end{cases}\]
Expand $b_i$\ w.r.t the root space decomposition:\ $b_i=b_i^{\mathfrak{h}}+\sum_{\alpha \in \Delta}b_i^{\alpha}.$\ Then   
for $\varphi \in C_c^{\infty}(M,\mathfrak{h}),\ \supp{\varphi}\subset \dom,$
\eqa{(A_i-I)d\varphi &=[\varphi,b_i]=[\varphi,b_i^{\mathfrak{h}}]+\sum_{\alpha \in \Delta}i\naiseki{\alpha}{\varphi}b_i^{\alpha}\\
&=i\sum_{\alpha \in \Delta}\naiseki{\alpha}{\varphi}b_i^{\alpha}.
}
Since $(A_i-I)d\varphi \in C_c^{\infty}(M,\mathfrak{h}),$\ it follows that $\naiseki{\alpha}{\varphi}b_i^{\alpha}=0.$\ Since this holds for any $\varphi$\ with $\supp{\varphi}\subset \dom,$\ we have $\text{Int}(\supp{b_i^{\alpha}}\cap \dom)=\phi.$\ 
Taking all the possible Cartan subalgebras as usual, we conclude that $\text{Int}(\supp{b_i}\cap \dom )=\phi\ .$\ From the above arguments,\ we find that for $\varphi \in C_c^{\infty}(M,\mathfrak{g}),\ \text{supp}(\varphi)\subset \mathcal{O},$\
\[(A_i-I)d\varphi =[\varphi,b_i]=0,\]
or equivalently,
\[A_id\varphi =d\varphi .\]
Therefore for $\psi \in N_0(\mathcal{O}),$\ we have
\[A_iV(\psi)d\varphi =V(\psi)A_id\varphi =V(\psi)d\varphi.\]
From Lemma\ \ref{totality}, we have only to prove  
\[A_iV(\psi)d\varphi =V(\psi)d\varphi \ \ \textbf{\underline{for all}}\ \psi \ \text{with}\ \supp{\psi}\subset \mathcal{O}.\tag{$\heartsuit $}\]
for the proof of $A_i|_{H(\mathcal{O})}=\text{Id}_{H(\mathcal{O})}$
(we have already shown that this ($\heartsuit $)\ is true for $\psi \in N_0(\mathcal{O})$).\ From Lemma\ \ref{separation}, there exists an open covering $\{V_k\}_{k=1}^N$\ of $K:=\supp{\psi}$\ and two families of functions $\{\psi_j^k\}_{0\le j\le n_k<\infty,\ 1\le k\le N}\subset N_0(\mathcal{O}),\ \{\varphi_k\}_{1\le k\le N}\subset \Mg,\ \supp{\varphi_k}\subset V_k$\ such that
\[
\begin{cases}
\dis \psi|_{V_k}&=\psi_{n_k}^k\psi_{n_k-1}^k\cdots \psi_{0}^k|_{V_k},\ \forall k,\\
d\varphi &= \dis \sum_{k=1}^Nd\varphi_k\ \text{on}\ K.
\end{cases}\] 
Thus, we obtain
\eqa{
A_iV(\psi)d\varphi &=\sum_{k=1}^NA_iV(\psi)d\varphi_k\\
&=\sum_{k=1}^NA_iV(\psi_l^k)\cdots V(\psi_0^k)d\varphi_k\\
&=\sum_{k=1}^NV(\psi_l^k)\cdots V(\psi_0^k)A_id\varphi_k\\
&=\sum_{k=1}^NV(\psi_l^k)\cdots V(\psi_0^k)d\varphi_k\\
&=V(\psi)d\varphi,
}
where in the third equality we used the equality $A_iV(\psi)=V(\psi)A_i\ \text{for}\ \psi\in N_0(\mathcal{O}).$\ 
Therefore $A_i|_{H(\dom)}=I_{H(\dom)}.$\ Furthermore, if $A_iH(\dom')\nsubseteq H(\dom'),$ there exists some $\omega \in H(\dom')$\ such that $\supp{A_i\omega}\cap \dom \neq \phi.$\ Again by the semisimplicity of $\mathfrak{g}$, there exists some $\psi \in N_0(\dom)$\ such that $V(\psi)A_i\omega \neq A_i\omega.$\ Since $V(\psi)\omega =\omega,$ it leads to a contradiction:
\[A_i\omega=A_iV(\psi)\omega=V(\psi)A_i\omega \neq A_i\omega.\]
Therefore we have $A_iH(\dom')\subset H(\dom')$ and we obtain $A_i\in \mathcal{A}(\dom').$\ The opposite inclusion $\mathcal{A}(\dom')\subset \mathcal{M}(\dom)'\cap \mathcal{A}$\ is obvious by the definition.\ Note that the equality $U(N_0)'\cap \mathcal{A}=\mathbb{C}1$ is also proved in the previous argument.\\
(2)\ The validity is obvious of the properties : isotony, locality\ (since $\supp{\psi_1}\cap \supp{\psi_2}=\phi \Rightarrow  \psi_1\psi_2=\psi_2\psi_1$) and additivity.\ This is an analogous situation to the case when there is an underlying Wightman field theory whose field operators $\phi (f)$\ are affiliated with the local algebras.\ In such a case additivity of the local net is always guaranteed.\ Furthermore, if we add the spectrum condition, as is usually assumed, cyclicity and separating property for the vacuum vector $\Omega $ is automatic\ (Reeh-Schlieder theorem\ \cite{Haag}).\ In order to examine the cyclicity and separating properties for our representation, consider the proper open subset $\mathcal{O}\subset M$.\ Make an orthogonal decomposition of the Hilbert space according to the support properties:
\[H\cong H(\mathcal{O})\oplus H(\mathcal{O}'),\ H(\mathcal{O}):=\{\omega \in H;\supp{\omega}\subset \mathcal{O}\}.\]
From the decomposition, we may use the identification which is an isometric isomorphism:
\eqa{
\Gamma(H)&\cong \Gamma(H(\mathcal{O}))\otimes \Gamma(H(\mathcal{O}')),\\
\EX{\omega_{\mathcal{O}}+\omega_{\mathcal{O}'}}&\leftrightarrow \EX{\omega_{\mathcal{O}}}\otimes \EX{\omega_{\mathcal{O}'}}.
}
Under the identification, we can compute the action\ of $\psi,\ \supp{\psi}\subset \mathcal{O}$:
\eqa{
U(\psi)\EX{\omega_{\dom}+\omega_{\dom'}}&=e^{-\frac{1}{2}||\beta(\psi)||^2-\naiseki{V(\psi)(\omega_{\dom}+\omega_{\dom'})}{\beta(\psi)}}\EX{\omega_{\mathcal{O}}+\omega_{\mathcal{O}'}}\\
&=e^{-\frac{1}{2}||\beta(\psi)||^2-\naiseki{V(\psi)\omega_{\dom}}{\beta(\psi)}}\EX{\omega_{\mathcal{O}}+\omega_{\mathcal{O}'}}\\ &=(U(\psi)|_{H(\dom )}\otimes I_{H(\dom')})(\EX{\omega_{\dom}}\otimes \EX{\omega_{\dom'}}).
}
Note that $\naiseki{V(\psi)\omega_{\dom'}}{\beta(\psi)}=0.$\ Therefore the local algebras have the following form:
\eqa{
\mathcal{M}(\dom)''&\cong \{U(\psi)|_{\Gamma(H(\dom))}\otimes I_{\Gamma(H(\dom'))};\supp{\psi}\subset \dom \}''\subset \mathbb{B}(\Gamma (H(\dom)))\otimes \mathbb{C}I_{\Gamma(H(\dom'))},\\
\mathcal{M}(\dom)'&\cong \{U(\psi)|_{\Gamma(H(\dom))};\supp{\psi}\subset \dom\}'\otimes \mathbb{B}(\Gamma (H(\dom'))).
}
From these forms it is clear that $\EX{0}$\ is not cyclic for $H$\ if $H(\dom')\neq \{0\}\ \blacksquare$.\\ \\
\textbf{Remark.}\\
\ \ It is clear that if the representation is irreducible, then $\Omega$ is not separating for $\mathcal{M}(\mathcal{O})''\ (\dom \neq \phi,M)$\ and any local algebras $\mathcal{M}(\dom)''$ are type I factors.\\
From the proof of (2), we see that Haag-type duality\footnote{This condition is crucial for the Doplicher-Haag-Roberts sector theory\ \cite{sector}.}\ $\mathcal{M}(\dom)''=\mathcal{M}(\dom')'$\ is equivalent to the following two conditions:
\[\begin{cases}
\{U(\psi)|_{\Gamma(H(\dom))};\supp{\psi}\subset \dom \}''=\mathbb{B}(\Gamma (H(\dom)))\\
\{U(f)|_{\Gamma(H(\dom'))};\supp{f}\subset \dom'\}''=\mathbb{B}(\Gamma(H(\dom')))
\end{cases}\]
We are not sure if the boundary behavior of the derivative of $\psi$ affects the irreducibility of the representation. Therefore it seems that the proof of the irreducibility for $\{U(f)|_{\Gamma(H(\dom'))};\supp{f}\subset \dom'\}$ requires more discussions, even if we have proved the irreducibility for the same dimensional manifolds.\ 
Finally, if we want to prove the irreducibility from our theorem, there is a difficulty concerning the strong limit.\ Let $\Xi \in \mathcal{M}'.$\ Then from the strong density of $\mathcal{A}$, there is a net $\{\Xi_{\alpha}\}\subset \mathcal{A}$, $\dis s-\lim \Xi_{\alpha}=\Xi.$\ We know from the proof of the above theorem that if for any $\alpha$, there exists some $\alpha>0$ such that $\Xi_{\alpha_0}$ commutes with all $U(\psi)$ where $\psi$\ belongs to some small neighborhood of 1, then $\Xi_{\alpha_0}=\lambda_{\alpha_0}I$. Taking subnet, we see that $\Xi$ is also a scalar operator.\ However, if for any $\alpha$, there is an operator $U(\psi)(\psi \in N)$ which does not commute with $\Xi_{\alpha}$ for any small neighborhood $N$ of $1\in \MG$, the situation is more subtle.\ Let $\{N_k\}_{k=1}^{\infty}$ be a family of neighborhoods of $1$ such that 
\[\{1\}\subset \cdots \subset N_{k+1}\subset N_k\subset \cdots \subset N_1,\ \CAP_{k=1}^{\infty}N_k=\{1\}.\]
The situation is as follows.\ For any $\alpha$ and any $k$, there exists some $\psi_{\alpha,k}\in N_k$ such that $\Xi_{\alpha}U(\psi_{\alpha,k})\neq U(\psi_{\alpha,k})\Xi_{\alpha}$, which implies 
\[\Xi_{\alpha}U(\psi_{\alpha,k})\EX{\omega_{\alpha,k}}\neq U(\psi_{\alpha,k})\Xi_{\alpha}\EX{\omega_{\alpha,k}},\ \exists \ \omega_{\alpha,k}\in H.\]
However, $\psi_{\alpha,k}\in N_k$ implies $\lim_{k\to \infty}\psi_{\alpha,k}=1$ and therefore we have
\[\dis s-\lim_{k\to \infty}U(\psi_{\alpha,k})=I.\]
Furthermore, by the assumption
\[\begin{cases}\dis s-\lim_{\alpha}\Xi_{\alpha}=\Xi ,\\
U(\psi)\Xi = \Xi U(\psi),\ \forall \psi \in \MG.
\end{cases}\]
Therefore to prove the irreducibility, we must derive some contradictions from these conditions, which looks quite non-trivial. The difficulty in this approach seems
to be different from those appearing in the Gaussian measure
analysis of the preceding researches in $\dim (M) = 2$\ \cite{Gel80,Alb81,Wal}.\ 
While our approach is formulated in a way independent of
the dimensionality of the manifold, such a possibility may not
be negated that the dimensionality might show up at certain
point in the process of taking suitable limits.\\ \\
\textbf{Acknowledgement}\\
The author would like to express his sincere thanks to Professor I. Ojima for his important advices, discussions and warm encouragements.\ He is also grateful to Mr. T. Hasebe, Mr. R. Harada Mr. K. Okamura and Mr. H. Saigo for invaluable discussions and comments. 

\end{document}